\documentclass[a4paper,10pt]{article}
\usepackage[utf8]{inputenc}
\usepackage[T1]{fontenc}
\usepackage{amsmath,amssymb,amsthm}
\usepackage{enumitem}
\usepackage[english]{babel}
\usepackage{graphicx}
\usepackage{listings}
\usepackage{dsfont}
\usepackage{geometry}
\usepackage{fancyvrb}
\usepackage{hyperref}
\begin{document}

\title{SMAGEXP: a galaxy tool suite for transcriptomics data meta-analysis}
\author{Samuel Blanck\,$^{1,}$\footnote{to whom correspondence should be addressed}~,  Guillemette Marot\,$^{1,2}$}

\maketitle

\begin{center}
{$^{1}$Univ. Lille Droit et Santé, EA 2694, F-59000 Lille, France \\
$^{2}$ Inria Lille-Nord Europe, MODAL, F-59000 Lille, France\\}
\end{center}

\begin{abstract}

\textbf{Background:}
With the proliferation of available microarray and high throughput sequencing experiments in the public domain, the use of meta-analysis methods increases. In these experiments, where the sample size is often limited, meta-analysis offers the possibility to considerably enhance the statistical power and give more accurate results. For those purposes, it combines either effect sizes or results of single studies in a appropriate manner. 
R packages metaMA and metaRNASeq perform meta-analysis on microarray and NGS data, respectively. They are not interchangeable as they rely on statistical modeling specific to each technology.  

\textbf{Results:}
SMAGEXP (Statistical Meta-Analysis for Gene EXPression) integrates metaMA and metaRNAseq packages into Galaxy. We aim to propose a unified way to carry out meta-analysis of gene expression data, while taking care of their specificities.
We have developed this tool suite to analyse microarray data from Gene Expression Omnibus (GEO) database or custom data from affymetrix$^{\mbox{\scriptsize{\copyright}}}$ microarrays. These data are then combined to carry out meta-analysis using metaMA package. SMAGEXP also offers to combine raw read counts from Next Generation Sequencing (NGS) experiments using DESeq2 and metaRNASeq package. In both cases, key values, independent from the technology type, are reported to judge the quality of the meta-analysis. These tools are available on the Galaxy main tool shed. Source code, help and installation instructions are available on github.

\textbf{Conclusion:} The use of Galaxy offers an easy-to-use gene expression meta-analysis tool suite based on the metaMA and metaRNASeq packages.

\end{abstract}


\section*{Background}

Meta-analyses are widely used in medicine and health policy to increase statistical power in studies suffering from small sample sizes. Gene expression experiments are a typical example of such designs. The R packages metaMA and metaRNASeq are dedicated to gene expression microarray and NGS meta-analysis, respectively.
While metaMA and metaRNASeq are open source and available on CRAN, they require coding skills in R to perform meta-analysis. Thus, to facilitate the use and the dissemination of these packages, we developed Galaxy wrappers.
Galaxy \cite{goecks2010,blankenberg2010,giardine2005} is an open, web-based platform for data intensive biomedical research. It keeps tracks of history and all analyses can be rerun. Galaxy community is very active and a lot of bioinformatics tools are included in Galaxy thanks to a modular system based on XML wrappers. These integrated tools can be shared via the Galaxy toolshed which serves as an appstore.%

\section*{Methods}

\subsection*{Overview of R packages integrated into Galaxy}

\subsubsection*{metaMA}
Gene expression microarray data meta-analysis can be performed thanks to the metaMA \cite{marot2009}  R package. It proposes methods to combine either p-values or moderated effect sizes from different studies to find differentially expressed genes. In our pipeline we only keep the inverse normal method \cite{Hedges1985} to combine the p-values calculated by limma \cite{Ritchie2015} for each single study.

\subsubsection*{metaRNAseq}
RNA-seq data meta-analysis can be performed thanks to the metaRNASeq \cite{Rau2014} R package. It implements two p-value combination techniques : the inverse normal and Fisher methods \cite{Fisher1932}. Single study p-values are comptuted with DESeq2 \cite{Love2014}. 

\subsubsection*{Differences between metaMA and metaRNASeq}

Main differences come from the statistical distributions used to model data and from the manner to treat the genes exhibiting conflicting expression patterns (i.e., under-expression when comparing one condition to another in one study, and over-expression for the same comparison in another study). Usually, microarray data are modelled by Gaussian distributions while NGS data are modelled by Negative Binomial distributions.
As explained in \cite{marot2009} and \cite{Rau2014}, the trick which consists to use one-tailed p-values for each single study before combination in metaMA avoids directional conflicts. In metaRNASeq, this trick can not be used, which necessits a post-hoc identification of conflicts, step which is also proposed in metaRNASeq.

\subsection*{Description of Galaxy tools}
\begin{figure*}[!h]
    \centering
	\includegraphics[width=1\textwidth]{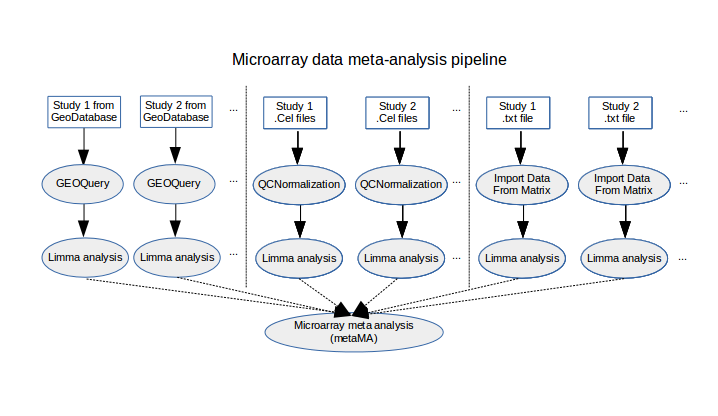} \\
  \caption{Overview of the tools from microarray data meta-analysis pipeline integrated within Galaxy.}
  \label{fig:pipeline}
\end{figure*}

SMAGEXP tool suite offers two distinct gene expression meta-analysis functionalities : one dedicated to microarray data meta-analysis and one dedicated to RNAseq data meta-analysis (see Table \ref{table:tools} and figure \ref{fig:pipeline}). 

\begin{table}[!h]
\caption{Summary of tools inputs and outputs.}
\label{table:tools}
	  \scalebox{0.75}{
      \begin{tabular}{l|l|l}
        \hline
        Tool & Input  &Output  \\ \hline
        GEOQuery & GEO database ID & Rdata object and .cond file \\
        QCNormalization & Raw .CEL affymetrix files and .conf file & Rdata object and plots \\
        Import custom data & Expression data in tabular .txt format  &Rdata object and plots  \\ 
        Limma analysis & Rdata object from GEOQuery or QCNormalization   & Rdata Object and HTML report \\
     	Microarray data meta-analysis & Rdata objects from Limma analyse  & HTML report  \\       
        RNA-seq data meta-analysis & Results text files from galaxy deseq2 tool  & HTML report  \\ \hline
      \end{tabular}
      }
\end{table}

\subsubsection*{Microarray data meta-analysis}

\paragraph*{GEOQuery tool}

GEOQuery tool fetches microarray data directly from GEO database, based on the GEOQuery \cite{Davis2007} R package. Given a GSE accession ID, it returns an Rdata object containing the data and a text file (.cond file) summarizing the conditions of the experiment. The .cond file is a text file containing one line per sample in the experiment. Each line is made of 3 columns:
\begin{itemize}[topsep=0pt,itemsep=-1ex,partopsep=1ex,parsep=1ex]
	\item Sample ID 
	\item Condition of the biological sample
	\item Description of the biological sample
\end{itemize}

\begin{figure*}[!ht]
    \centering
	\includegraphics[width=0.5\textwidth]{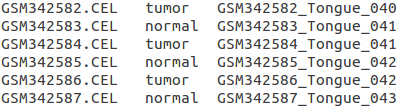} \\
  \caption{Example of .cond file.}
  \label{figure:condfile}
\end{figure*}  

Column names are optional and only the columns order matters. 
As the GEO dataset should already have been normalized, the GEOQuery tool does not perform any normalization method, apart from an optional log2 transformation.


\paragraph*{QCNormalization tool}
It is possible to analyze .CEL files from affymetrix gene expression microarray. The QCnormalization tool offers to ensure the quality of the data and to normalize them. Several normalization methods are available :
\begin{itemize}[topsep=0pt,itemsep=-1ex,partopsep=1ex,parsep=1ex]
	\item rma normalization 
	\item quantile normalization + log2
	\item background correction + log2
	\item log2 only
\end{itemize}

This tool generates several quality figures : microarray images, boxplots and MA plots. It also outputs an Rdata object containing the normalized data for further analysis with the limma analysis tool.

\paragraph*{Import custom data tool}
This tool imports data stored in a tabular text file. Column titles (chip IDs) must match the IDs of the .cond file. A few  normalization methods are proposed, but it is possible to skip the normalization step, by choosing "none" in the normalization methods options. Therefore this tool is of special interest when the input dataset has been previously normalized.

This tool also generates boxplots and MA plots and  outputs an Rdata object containing the data for further analysis with the limma analysis tool.


\paragraph*{Limma analysis tool}

\begin{figure*}[!ht]
    \centering
	\includegraphics[width=1\textwidth]{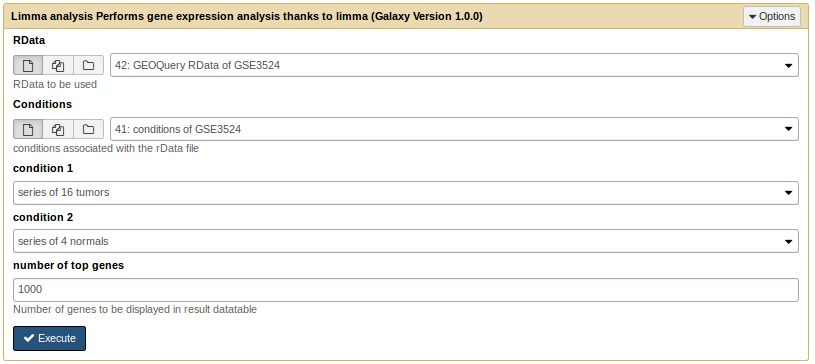} \\
  \caption{limma analysis tool form.}
  \label{figure:limmaForm}
\end{figure*}

The Limma analysis tool performs single analysis either of data previously retrieved from GEO database or normalized affymetrix .CEL files data. 
Given a .cond file, it runs a standard limma differential expression analysis. The user choose two conditions extracted from the .cond file (see Figure \ref{figure:limmaForm}). It generates boxplots for rough quality control of normalization, p-value histograms to ensure that statistical hypotheses are not violated and a volcano plot to quickly identify the most-meaningful changes. This tool also outputs a table summarizing the differentially expressed genes and their annotations. Genes are sorted by ascending Benjamini-Hochberg adjusted p-value, and annotations are retrieved via GEO database. This list of genes can be exported to excel or to csv format. This table is sortable and requestable. Furthermore it is possible to expand each row to display extended annotations informations, including hypertext links to the National Center for Biotechnology Information (NCBI) gene database.
Finally, this tool outputs an Rdata object to perform further meta-analysis and a tabular file containing the all results and annotations of the differential analysis. 

\paragraph*{Microarray data meta-analysis tool}
The meta-analysis relies on the metaMA R package. Prior to the meta-analysis itself, a pre-processing is made in order to ensure compatiblity between several sources of data. In fact, data could come from different types of microarrays. First, we list the Entrez gene ID corresponding to each probe of each dataset. Next, we keep the probes corresponding to the genes which are shared by all the experiments of the meta-analysis. Then, for each dataset, we merge the microarray probes originating from the same Entrez gene ID by computing their mean. Note that the merging of different technologies induces a loss of information and might generate several conflicts as probes do not necessary reflect the same biological reality.
Finally, the p-value combination method of metaMA is run on the merged dataset. It generates a Venn Diagram summarizing the results of the meta-analysis, and a list of indicators to evaluate the quality of the performance of the meta-analysis :
\begin{itemize}[topsep=0pt,itemsep=-1ex,partopsep=1ex,parsep=1ex]
	\label{item:indicators}
	\item DE : Number of differentially expressed genes 
	\item IDD (Integration Driven discoveries) : number of genes that are declared differentially expressed in the meta-analysis that were not identified in any of the single studies alone
	\item Loss : Number of genes that are identified differentially expressed in single studies but not in meta-analysis 
	\item IDR (Integration-driven Discovery Rate) : corresponding proportion of IDD
	\item IRR (Integration-driven Revision) : corresponding proportion of Loss
\end{itemize}
It also outputs a fully sortable and requestable table, with gene annotations and hypertext links to NCBI gene database.

\subsubsection*{RNA-seq data meta-analysis}

The RNA-seq data meta-analysis tool relies on the deseq2 galaxy tool analysis results. Given several text file resulting from the deseq2 tool, the metaRNAseq tool performs a meta-analysis, generates the list of differentially expressed genes, and outputs the DE, IDD, Loss, IDR and IRR indicators.

\section*{Application}

\subsection*{Microarray meta-analysis example}
\begin{figure*}[!ht]
    \centering
	\includegraphics[width=1\textwidth]{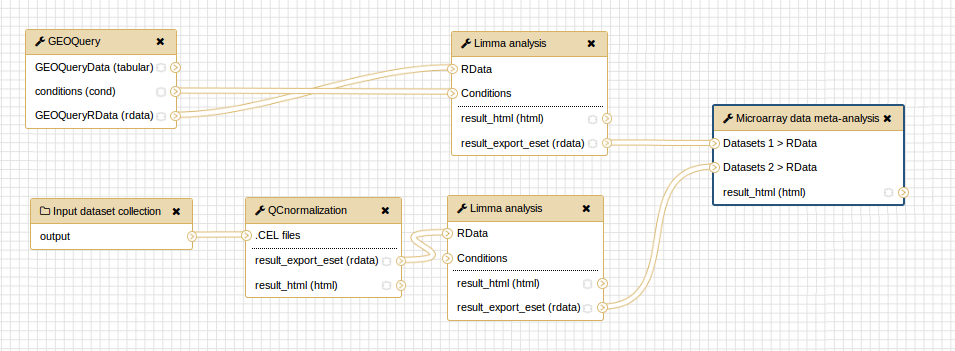} \\
  \caption{Exemple of a galaxy workflow for microarray meta-analysis.}
  \label{figure:fig2}
\end{figure*}

SMAGEXP was applied to two GEO datasets identified with the following IDs : GSE3524 and GSE13601. These two datasets contain  human oral squamous cell carcinoma (SCC) data. See Figure \ref{figure:fig2} for an overview of the worfklow of this analysis.

\begin{figure*}[!ht]
    \centering
	\includegraphics[width=0.70\textwidth]{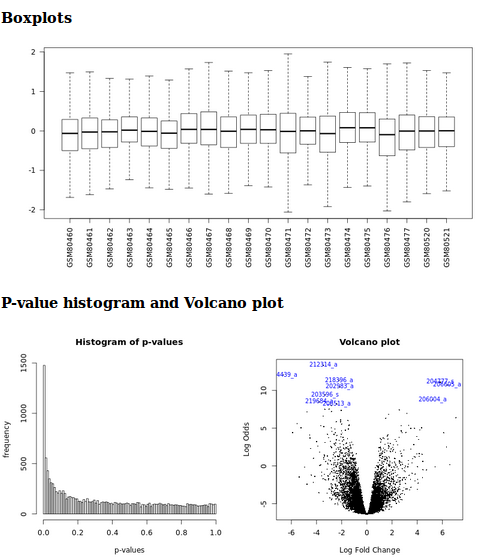} \\
  \caption{limma analysis tool output plots.}
  \label{figure:limmaplots}
\end{figure*}

\begin{figure*}[!ht]
    \centering
	\includegraphics[width=1\textwidth]{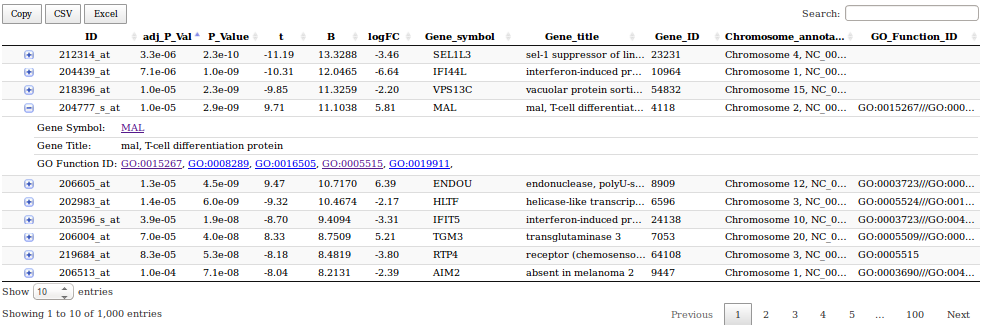} \\
  \caption{limma analysis tool : table of top 10 genes for GSE3524 dataset.}
  \label{figure:limmatable}
\end{figure*}

First, we fetch data from the GSE3524 using the GEOQuery tool (with parameter "log2 transformation" = auto). Then we launch the limma analysis, using the output from the GEOquery tool. It generates an Rdata output, which will be usefull for the meta-analysis.
Results can be seen on Figure \ref{figure:limmaplots} and Figure \ref{figure:limmatable}

Secondly, the same kind of analysis is run from raw .CEL files. We choose to keep six .CEL files  from the GSE13601 dataset (IDs from GSM342582 to GSM342587).
Quality control and normalization is done thanks to the QCnormalization tool. 
Then, as previously, the limma analysis tool is run to generate a HTML report and an Rdata output.

\subsubsection*{Run a metaMA analysis}
\begin{figure*}[h!]
    \centering
	\includegraphics[width=1\textwidth]{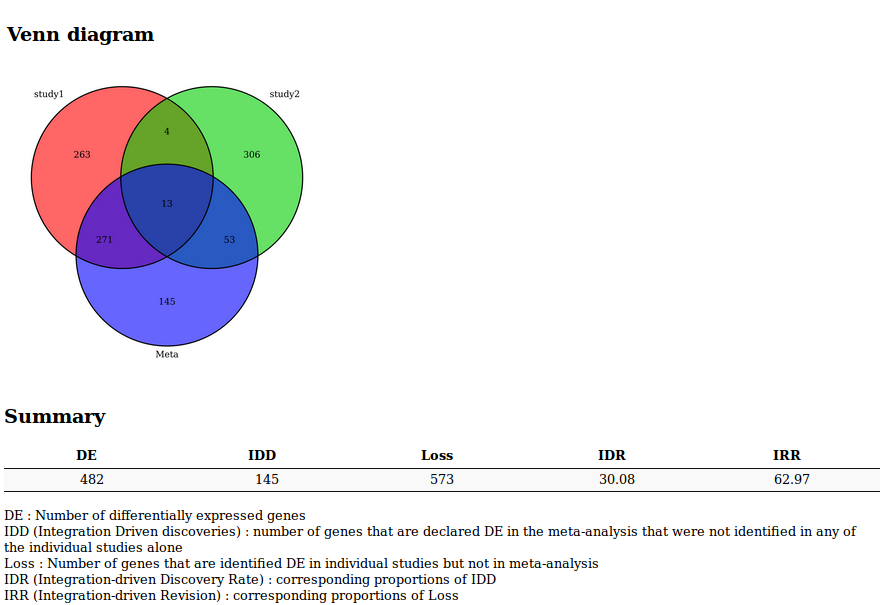} \\
  \caption{Venn diagram and summary of microarray data meta-analysis tool results}
  \label{figure:metaMAsummary}
\end{figure*}

To run the microarray meta-analysis tool, we only need the Rdata output of each single study, generated by the limma analysis tool. It generates a Venn diagram to compare the results of each study with the meta-analysis. It also outputs several indicators as described in the description of the tool (see Figure \ref{figure:metaMAsummary}). As for the limma tool, annotated expressed genes are displayed in a table which can be ordered and requested.

\subsection*{RNA-seq data meta-analysis example}
\begin{figure*}[h!]
    \centering
	\includegraphics[width=1\textwidth]{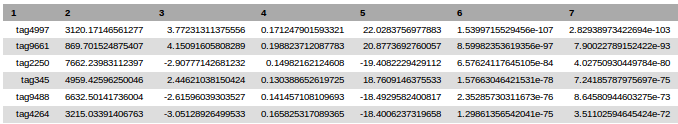} \\
  \caption{Header of a deseq2 results file}
  \label{figure:deseq2results}
\end{figure*}

The RNA-seq data meta-analysis tool relies on deseq2 results (see Figure \ref{figure:deseq2results}). 

It outputs a Venn diagram and the same indicators as in the microarray data analysis tool for both Fisher and inverse normal p-values combinations.
It also generates a text file containing summarization of the results of each single analysis and meta-analysis. Potential conflicts between single analyses are indicated by zero values in the "signFC" column.

\begin{figure*}[h!]
    \centering
	\includegraphics[width=1\textwidth]{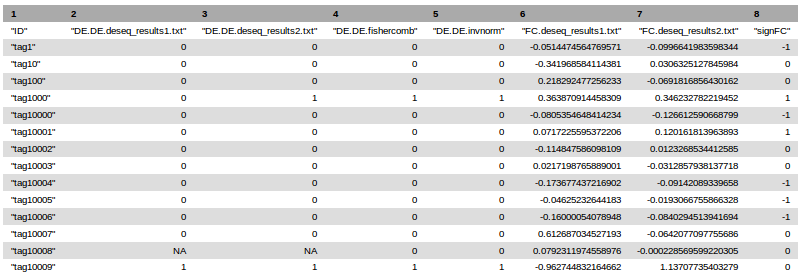} \\
  \caption{Header of a metaRNAseq results file}
  \label{figure:metaRNAresults}
\end{figure*}

\section*{Conclusion}
We developed SMAGEXP, a toolsuite dedicated to gene-expression data meta-analysis. This toolsuite proposes quality controls, single analyses and meta-analyses of microarray and RNA-seq data, suggesting appropriate pipelines for each type of data. It delivers fully annnotated results of differentially expressed genes, exportable in several usual formats. Integrated into Galaxy, SMAGEXP is easy to use for biologists and life scientists. R packages metaMA and metaRNAseq thus inherit reproductibility and accessibility support from Galaxy.
SMAGEXP is available on the Galaxy main toolshed \cite{Blankenberg2014}. \\ Source code is available on github at : https://github.com/sblanck/smagexp. \\ Furthermore, thanks to Docker, we made these Galaxy tools and their dependencies easy to deploy. A fully dockerized instance of Galaxy containing SMAGEXP is available at : \\ https://hub.docker.com/r/sblanck/galaxy-SMAGEXP/


\bibliographystyle{ieeetr}
\bibliography{arxiv_article}

\end{document}